# Spin torque nano-oscillators based on antiferromagnetic skyrmions

Laichuan Shen [1, 2, *], Jing Xia [2, *], Guoping Zhao [1, †], Xichao Zhang [2, 3], Motohiko Ezawa [4], Oleg A. Tretiakov [5, 6, 7], Xiaoxi Liu [3], and Yan Zhou [2, †]

[1]*College of Physics and Electronic Engineering, Sichuan Normal University, Chengdu 610068, China*

[2]*School of Science and Engineering, The Chinese University of Hong Kong, Shenzhen, Guangdong 518172, China*

[3]*Department of Electrical and Computer Engineering, Shinshu University, 4-17-1 Wakasato, Nagano 380-8553, Japan*

[4]*Department of Applied Physics, The University of Tokyo, 7-3-1 Hongo, Tokyo 113-8656, Japan*

[5]*Institute for Materials Research and Center for Science and Innovation in Spintronics, Tohoku University, Sendai 980-8577, Japan*

[6]*School of Physics, The University of New South Wales, Sydney 2052, Australia*

[7]*National University of Science and Technology "MISiS", Moscow 119049, Russia*

\* These authors contributed equally to this work.

† Authors to whom correspondence should be addressed:

E-mail (G.Z.): zhaogp@uestc.edu.cn & E-mail (Y.Z.): zhouyan@cuhk.edu.cn

**Abstract**

Skyrmion-based spin torque nano-oscillators are potential next-generation microwave signal generators. However, ferromagnetic skyrmion-based spin torque nano-oscillators cannot reach high oscillation frequencies. In this work, we propose to use the circular motion of an antiferromagnetic skyrmion to create the oscillation signal in order to overcome this obstacle. Micromagnetic simulations demonstrate that the antiferromagnetic skyrmion-based spin torque nano-oscillators can produce high frequencies (tens of GHz). Furthermore, the speed of the circular motion for an antiferromagnetic skyrmion in a nanodisk is analytically derived, which agrees well with the results of numerical simulations. Our findings are useful for the understanding of the inertial dynamics of an antiferromagnetic skyrmion and the development of future skyrmion-based spin torque nano-oscillators.

Since spin-transfer torque (STT) was predicted [1,2], it has received great attention due to its critical role in modern magnetic devices, such as spin-transfer torque magnetic random access memory (STT-MRAM) [3,4] and spin torque nano-oscillators (STNO) [5-17]. In particular, STNO can be used as microwave signal generator, where microwave signal is generated by the precession of the uniform magnetization [10] or gyrotropic motion of a magnetic vortex [7,11,16]. Recently, numerical calculations show that the oscillations of magnetic skyrmions can also excite microwave signals with small line width and are expected to improve the output power [12,18-21]. Magnetic skyrmions are swirling spin textures, which have topologically protected stability and low depinning current [22,23]. In most systems, such as in $Fe_{1-x}Co_xSi$ [24], Ir/Co/Pt [25], Pt/CoFeB/MgO [26], Co/Pd [27] and Pt/Co/MgO [28], magnetic skyrmions are stabilized by the Dzyaloshinskii-Moriya interaction (DMI) [29,30], which is induced by the lack or breaking of inversion symmetry in magnetic materials. Compared to the vortex-based STNO, the skyrmion-based STNO has some advantages. For instance, multiple skyrmions can be put in one STNO to improve the oscillation frequency [12,19]. Generally, a higher frequency signal has a larger tuning range of frequency. And more data can be sent in less time for the higher-frequency transmissions. Besides, skyrmions are localized spin textures and could have a much smaller size than the magnetic vortices [16,31].

However, for the STNO based on the skyrmion motion in ferromagnetic (FM) nanodisks, the range of the oscillation frequency is narrow (it is from 0 Hz to about 1 GHz for the STNO with one FM skyrmion). [12,18-20] Therefore, it is necessary to explore alternative skyrmion-based STNO with broader frequencies. Compensated antiferromagnets are promising material systems for spintronic devices [32-36] because they have ultrafast magnetization dynamics and no stray magnetic fields. Theoretical calculations [37-39] show the existence of the stable skyrmions in antiferromagnets, and recently the stabilization of magnetic skyrmions has been experimentally presented in ferrimagnetic GdFeCo films, [40] which have similar spin structure to antiferromagnets. Compared to ferromagnetic skyrmions, antiferromagnetic (AFM) skyrmions have no skyrmion Hall effect [37,38,41], so that their motion trajectory is a perfect straight line along the driving force direction. Therefore, AFM skyrmions are ideal information carriers for racetrack-type memory [42-45]. For skyrmion-based STNO, however, the study of the differences between FM and AFM skyrmions is still lacking.

In this work, we analytically and numerically study the motion of a FM skyrmion and an AFM skyrmion in a nanodisk driven by spin-polarized currents with a vortex-like polarization [18,20]. It is found that since the physical mechanism of their steady motion is different, the oscillation frequency of AFM skyrmion-based STNO (tens of GHz) is reasonably higher than that of FM skyrmion-based STNO (~ 1 GHz) [12,18-20].

The model of the skyrmion-based STNO is depicted in Fig. 1. The fixed layer with a magnetic vortex configuration is used to generate the spin-polarized current with a vortex-like polarization. [18,20,46] Such a spin-polarized current applies spin torques to the local magnetic moments of the free layer and then the skyrmion in the nanodisk moves in a circular motion. Thus, using the skyrmion-based STNO, a direct current can induce an oscillating signal. The signal can be detected by constructing nano-contact oscillator and using the magnetoresistance effect [12,21]. Due to the fixed layer with a vortex magnetic configuration, the polarization vector *p* depends on the spatial coordinates (*x*, *y*) and is described as *p* = (cos$\beta$, sin$\beta$, 0) with $\beta$ = arctan(*y*/*x*) + $\varphi$. [18] In this work, unless otherwise noted, the polarized angle $\varphi$ is set at 0° [see Fig. 1(b)], i.e., taking the divergent vortex in the fixed layer, [18] where the divergent vortex exists in many systems, such as in Co/Rh/NiFe [47], NiFe/Cr/NiFe [48], NiO/Fe [49] and CoO/Fe [49]. For AFM skyrmion-based STNO, the influence of $\varphi$ on the oscillation frequency is shown in Fig. S1 of the supplementary material, whereas for FM skyrmion-based STNO, it has been reported [18].

Although the principle of AFM skyrmion-based STNO is similar to that of FM skyrmion-based STNO, the physical mechanism of their steady motions is different. Taking the current density *j* = ± 20 MA/cm$^2$ to drive Néel-type FM and AFM skyrmions in a nanodisk with radius *R* = 60 nm, Fig. 2 shows the simulated trajectories of the guiding center for the FM and AFM skyrmions. For comparison, the parameters used here are the same for both FM and AFM skyrmions (details of the simulations are given in supplementary material). The guiding center of AFM skyrmion is defined as

$$r_i = \frac{\int dxdy[i\mathbf{n}\cdot(\partial_x\mathbf{n}\times\partial_y\mathbf{n})]}{\int dxdy[\mathbf{n}\cdot(\partial_x\mathbf{n}\times\partial_y\mathbf{n})]}, \quad i = x, y, \qquad (1)$$

while for FM skyrmion, we need to replace the AFM Néel vector *n* with the FM reduced magnetization *m* (*m* = *M*/*M*$_S$ with the saturation magnetization *M*$_S$). As shown in Fig. 2, the AFM skyrmion moves steadily in the nanodisk independently of the sign of the applied current. Meanwhile, the FM skyrmion moves toward the nanodisk center when the positive current is applied, whereas for the negative current, the skyrmion is destroyed at the nanodisk edge (see Movie 1). It should be noted that if the negative current is small, the FM skyrmion can also move steadily.

In order to analyze the dynamics of FM and AFM skyrmions, we derived Thiele equations [50-52]. Considering that a rigid FM skyrmion moves steady, from the Landau-Lifshitz-Gilbert (LLG) equation [41,53,54], we obtain the Thiele equation, [12,18,20,44,55-57]

$$\boldsymbol{G} \times \boldsymbol{v} + \boldsymbol{F}_\alpha + \boldsymbol{F}_j + \boldsymbol{F}_b = \boldsymbol{0}, \quad (2)$$

where the first term is the Magnus force and the gyrovector $\boldsymbol{G} = 4\pi Q \mu_0 M_S t_z/\gamma \boldsymbol{e}_z$ with the vacuum permeability constant $\mu_0$, the free layer thickness $t_z$ and the gyromagnetic ratio $\gamma$. The skyrmion number $Q = -\frac{1}{4\pi}\int dxdy [\boldsymbol{m} \cdot (\partial_x \boldsymbol{m} \times \partial_y \boldsymbol{m})]$ is equal to $\pm 1$ for an isolated FM skyrmion. [37,41,58] The second term of Eq. (2) stands for the dissipative force, $\boldsymbol{F}_\alpha = -\alpha \mu_0 M_S t_z \boldsymbol{v} \cdot \boldsymbol{d}/\gamma$ with the damping $\alpha$, the velocity $\boldsymbol{v}$ and the dissipative tensor $\boldsymbol{d} = \begin{pmatrix} d & 0 \\ 0 & d \end{pmatrix}$, where $d = \int dxdy\, \partial_x \boldsymbol{m} \cdot \partial_x \boldsymbol{m}$. The third term $\boldsymbol{F}_j$ represents the driving force induced by the current that can be decomposed into two orthogonal parts (the radial $\boldsymbol{e}_r$ and tangential $\boldsymbol{e}_t$ directions based on the symmetry of a nanodisk), i.e., $\boldsymbol{F}_j = F_{jt}\boldsymbol{e}_t + F_{jr}\boldsymbol{e}_r$ with $F_{ji} = -\mu_0 B_j M_S t_z \int dxdy [(\boldsymbol{m} \times \boldsymbol{p}) \cdot \partial_i \boldsymbol{m}]$, where $B_j$ relates to the applied current density $j$, defined as $j\hbar P/2\mu_0 e M_S t_z$ with the reduced Plank constant $\hbar$, the spin polarization rate $P$, and the elementary charge $e$. The last term is the boundary-induced force, $\boldsymbol{F}_b = -\nabla U$ with the potential energy $U$.

On the other hand, based on the spin dynamics equations for antiferromagnets, [59-64] the Thiele equation for AFM skyrmion is obtained, which is similar to Newton's kinetic equation, described as (see supplementary material for details), [38,60]

$$\boldsymbol{a} \cdot M_{\text{AFMSk}} = \boldsymbol{F}_\alpha + \boldsymbol{F}_j + \boldsymbol{F}_b, \quad (3)$$

where $\boldsymbol{a}$ is the acceleration, and $M_{\text{AFMSk}}$ is the effective AFM skyrmion mass which is defined as $\mu_0^2 M_S^2 t_z \boldsymbol{d}/2A_h\gamma^2$ with homogeneous exchange constant $A_h$. Similar to the case of the FM skyrmion, three terms on the right side of Eq. (3) stand for the dissipative, current-induced and boundary-induced forces, respectively. It should be noted that the involved physical quantities in Eqs. (2) and (3) are described by the FM reduced magnetization $\boldsymbol{m}$ and the AFM Néel vector $\boldsymbol{n}$ respectively.

Comparing Eqs. (2) and (3), we find two distinct differences between FM and AFM skyrmions. First, an AFM skyrmion consists of two FM skyrmions (one on each sublattice) with opposite skyrmion numbers for magnetization, i.e., $Q = 0$, so that the net Magnus force is zero [37,38]. Meanwhile, a FM skyrmion has skyrmion number $Q = \pm 1$, so that the Magnus force is always nonzero and leads to the skyrmion Hall effect [55,65]. Second, Eq. (3) shows that the AFM skyrmion has an effective mass $M_{\text{AFMSk}}$, which is inversely proportional to the homogeneous exchange constant $A_h$, while Eq. (2) does not describe the inertial dynamics [26,66].

The above two differences between FM and AFM skyrmions cause their different dynamic behaviors in the nanodisk. To show their difference clearly, Thiele equations for FM and AFM skyrmions are decomposed into two orthogonal parts (for the radial and tangential directions). When a skyrmion moves steadily in the nanodisk, there is only velocity $v_t$ in the tangential direction. For the case of the divergent polarization vector, $F_{jr} \sim 0$ (see supplementary material). Thus, Eq. (2) describing the FM skyrmion is split into

$$v_t \mathbf{G} \times \mathbf{e_t} - F_b \mathbf{e_r} = \mathbf{0}, \quad (4a)$$

$$F_\alpha \mathbf{e_t} + F_{jt} \mathbf{e_t} = \mathbf{0}. \quad (4b)$$

On the other hand, for the AFM skyrmion, Eq. (3) is decomposed as

$$a_r M_{\text{AFMSk}} \mathbf{e_r} = F_b \mathbf{e_r}, \quad (5a)$$

$$F_\alpha \mathbf{e_t} + F_{jt} \mathbf{e_t} = \mathbf{0}. \quad (5b)$$

By comparing Eqs. (4a) and (5a), it can be seen that the steady circular motion of a FM skyrmion in a nanodisk requires the boundary-induced force $F_b$ to balance the Magnus force, while it requires $F_b$ to act as the centripetal force for an AFM skyrmion (see Fig. S2 in supplementary material). For a FM skyrmion, when $j$ is positive, the boundary-induced and Magnus forces are pointing along the line going through the center of the nanodisk, so that the FM skyrmion moves toward the nanodisk center eventually. For the negative current, although the Magnus force points toward the edge of the nanodisk, the boundary-induced force is not strong enough to counterbalance the Magnus force, resulting in the destruction of the FM skyrmion at the nanodisk edge. To examine this result, we estimate the involved forces. Assuming motion radius $r_c$ of ~ 40 nm and speed $v_t$ of ~ 1 km/s, and combining the parameters used in this work, the Magnus force is estimated to be ~ $4 \times 10^{-11}$ N, which is about an order of magnitude larger than the boundary-induced force (~ $2 \times 10^{-12}$ N), see supplementary material. So, the FM skyrmion will be destroyed at the nanodisk edge. However, for the AFM skyrmion, the required centripetal force, $\frac{v_t^2}{r_c} M_{\text{AFMSk}}$, is ~ $2 \times 10^{-13}$ N based on the adopted parameters, so that the AFM skyrmion can move steadily in the nanodisk (see Movie 2). Besides, whether the sign of the applied current is positive or negative, the required centripetal force does not change for $\varphi = 0°$. Therefore, the usage of the AFM skyrmion allows to naturally overcome this impediment, where a FM skyrmion moves toward the nanodisk center [12].

For the case of Fig. 2(b), the guiding center and velocity are oscillating steadily, and their amplitudes give radius $r_c$ ~ 34.74 nm and speed $v_t$ ~ 1.822 km/s of the circular motion (see Fig. S3 in supplementary material). It is noteworthy that a relatively high oscillation frequency is obtained, $f = v_t/2\pi r_c$ = 8.35 GHz, which is larger than the reported results for the FM skyrmion-based STNO [12,18-20].

In order to test the reliability of the simulated speeds, we derived the tangential speed $v_t$ of the circular motion from Eq. (5b), which can be written as (see supplementary material),

$$v_t \approx \frac{\pi^2 \gamma B_j R_s}{\alpha d}, \qquad (6)$$

where $R_s$ is the skyrmion radius given by $R_s \approx \frac{\Delta}{\sqrt{2 - 2D/D_c}}$ [67]. Here $\Delta = (A/K)^{1/2}$ is the domain wall width parameter and $D_c = 4(AK)^{1/2}/\pi$. $A$, $K$ and $D$ stand for the inhomogeneous exchange constant, perpendicular magnetic anisotropy constant, and DMI constant, respectively. Substituting the numerical values of $R_s$ ~ 7 nm and $d$ ~ 15.38 into Eq. (6), we obtain speed $v_t \approx 1.794/\alpha$ m/s per MA/cm$^2$. Taking the current density $j$ = 20 MA/cm$^2$ and damping $\alpha$ = 0.02 used in Fig. 2, the analytical speed $v_t$ is 1.794 km/s, which is in line with the above simulated speed 1.822 km/s. Moreover, Eq. (6) shows that the larger skyrmion size, which can be caused by increasing $D$ or decreasing $K$, gives faster motion and further increases frequency $f$.

In addition, frequency $f$, speed $v_t$, and radius $r_c$ are calculated as functions of nanodisk radius $R$, damping constant $\alpha$, and applied current $j$, as shown in Fig. 3. In Fig. 3(a), one can see that $f$ increases with the reduction of the nanodisk size due to the decrease of motion radius $r_c$ [see Fig. 3(g)]. In particular, for $R$ = 30 nm, $f$ takes a large value of 25 GHz. On the other hand, as shown in Fig. 3(d), the speed becomes larger with increasing nanodisk radius and then saturates at ~ 3.55 km/s because the edge compresses the skyrmion size for a small nanodisk [67]. One can see from Fig. 3(b) that the small damping $\alpha$ is advantageous to obtain a high $f$. Fig. 3(e) shows that $v_t$ depends significantly on $\alpha$, as expected from Eq. (6), where $v_t$ is inversely proportional to $\alpha$. When $\alpha$ goes down so that $v_t$ increases, the required centripetal force increases, resulting in that the AFM skyrmion moves along the larger orbit with a stronger boundary-induced force to act as the centripetal force [see Fig. 3(h)]. Another convenient and important method to tune the frequency is to change the applied current $j$, as shown in Fig. 3(c). The frequency $f$ increases from 0 to ~ 13 GHz when the applied current $j$ changes from 0 to 25 MA/cm$^2$, since $v_t$ increases almost linearly with $j$ [see Figs. 3(f)]. In addition, speed $v_t$ in our numerical simulations agrees well with the result of Eq. (6) [see Figs. 3(e) and 3(f)]. It should be mentioned that there is a critical current $j_c$ of ~ 27 MA/cm$^2$ above which the AFM skyrmion is destroyed at the nanodisk edge. Eqs. (5a) and (6) indicate that $j_c$ depends on the boundary-induced force and is proportional to $(1/M_{\text{AFMSk}})^{1/2}$, as shown in Fig. S4 of the supplementary material.

In summary, we have proposed a novel spin torque nano-oscillator based on the AFM skyrmion, where the boundary-induced force acts as the centripetal force to maintain the AFM skyrmion in a circular motion. The numerical simulations show that the AFM skyrmion can move at speeds of a few kilometers per second and such a nano-oscillator gives relatively high frequencies (tens of GHz). Furthermore, we derived analytically the motion speed, which becomes larger with the increase of the applied current and skyrmion size or reduction of the damping, that is in good agreement with the numerical simulations. Our results provide a promising route to modulate the frequency in a wide range for future skyrmion-based nano-oscillators.

See supplementary material for the details of micromagnetic simulations and analytical derivations for FM and AFM skyrmions in a nanodisk. The additional Movie 1 shows the destruction of a FM skyrmion at the nanodisk edge, and Movie 2 presents the steady circular motion for an AFM skyrmion in a nanodisk.

# Acknowledgement

G.Z. acknowledges the support by the National Natural Science Foundation of China (Grant Nos. 51771127, 51571126 and 51772004) of China, the Scientific Research Fund of Sichuan Provincial Education Department (Grant Nos. 18TD0010 and 16CZ0006). X.Z. was supported by JSPS RONPAKU (Dissertation Ph.D.) Program. M.E. acknowledges the support by the Grants-in-Aid for Scientific Research from JSPS KAKENHI (Grant Nos. JP18H03676, JP17K05490 and JP15H05854), and also the support by CREST, JST (Grant Nos. JPMJCR16F1 and JPMJCR1874). O.A.T. acknowledges support by the Grants-in-Aid for Scientific Research (Grant Nos. 17K05511 and 17H05173) from MEXT, Japan, by the grant of the Center for Science and Innovation in Spintronics (Core Research Cluster), Tohoku University, by JSPS and RFBR under the Japan-Russian Research Cooperative Program, and by the Ministry of Education and Science of the Russian Federation in the framework of Increase Competitiveness Program of NUST "MISiS" (Grant No. K2-2017-005), implemented by a governmental decree dated $16^{th}$ of March 2013, N 211. X.L. acknowledges the support by the Grants-in-Aid for Scientific Research from JSPS KAKENHI (Grant Nos. 17K19074, 26600041 and 22360122). Y.Z. acknowledges the support by the President's Fund of CUHKSZ, the National Natural Science Foundation of China (Grant No. 11574137), and Shenzhen Fundamental Research Fund (Grant Nos. JCYJ20160331164412545 and JCYJ20170410171958839).

**Figure Captions**

Fig. 1. (a) The sketch of skyrmion-based spin torque nano-oscillators. The fixed layer with a vortex magnetic configuration is used to generate the spin-polarized current, where the angle between the polarization vector $p$ and the radial direction $e_r$ is defined as $\varphi$. The heavy-metal layer is necessary to induce the interfacial DMI, which can stabilize skyrmions. (b) The configurations of the polarization vector $p$ for $\varphi = 0°$ and $90°$. In our simulations, $1 \times 1 \times 1$ nm$^3$ is employed to discrete the free layer with thickness $t_z = 1$ nm. In addition, the following parameters are adopted: [66] $A = 15$ pJ/m, $K = 0.6$ MJ/m$^3$, $M_S = 580$ kA/m, $D = 3$ mJ/m$^2$, $\gamma = 2.211 \times 10^5$ m/(A s), $P = 0.4$, $\alpha = 0.01 \sim 0.2$ and $A_h = 10$ MJ/m$^3$.

Fig. 2. (a) The trajectory for an AFM skyrmion driven by a current ($j = 20$ MA/cm$^2$). (b) The trajectory for an AFM skyrmion driven by a current ($j = -20$ MA/cm$^2$). (c) The trajectory for a FM skyrmion driven by a current ($j = 20$ MA/cm$^2$). (d) The trajectory for a FM skyrmion driven by a current ($j = -20$ MA/cm$^2$). For comparison, the adopted parameters are the same for both FM and AFM skyrmions, $\alpha = 0.02$ and $R = 60$ nm. Solid lines represent the trajectories and dashed lines stand for the nanodisk edges.

Fig. 3. The influences of nanodisk radius $R$, damping constant $\alpha$ and current density $j$ on frequency $f$ [(a)-(c)], speed $v_t$ [(d)-(f)] and radius $r_c$ [(g)-(i)] for AFM skyrmion-based nano-oscillators. Symbols stand for the numerical simulation results and dash lines are guiding the eyes. The solid lines are the analytical speeds obtained from Eq. (6). In (a), (d) and (g), $\alpha = 0.01$ and $j = 20$ MA/cm$^2$. In (b), (e) and (h), $R = 60$ nm and $j = 20$ MA/cm$^2$. In (c), (f) and (i), $R = 60$ nm and $\alpha = 0.01$.

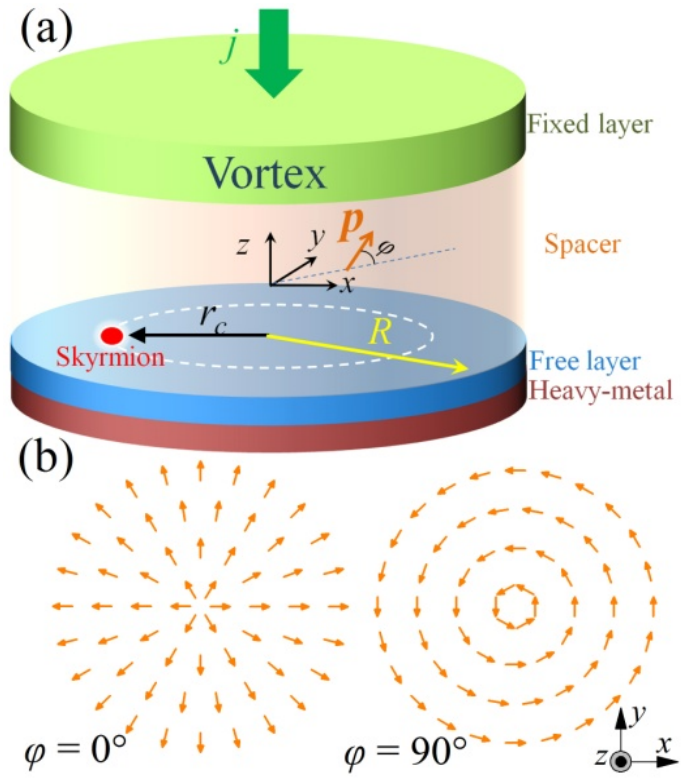

Figure 1

Shen et al.

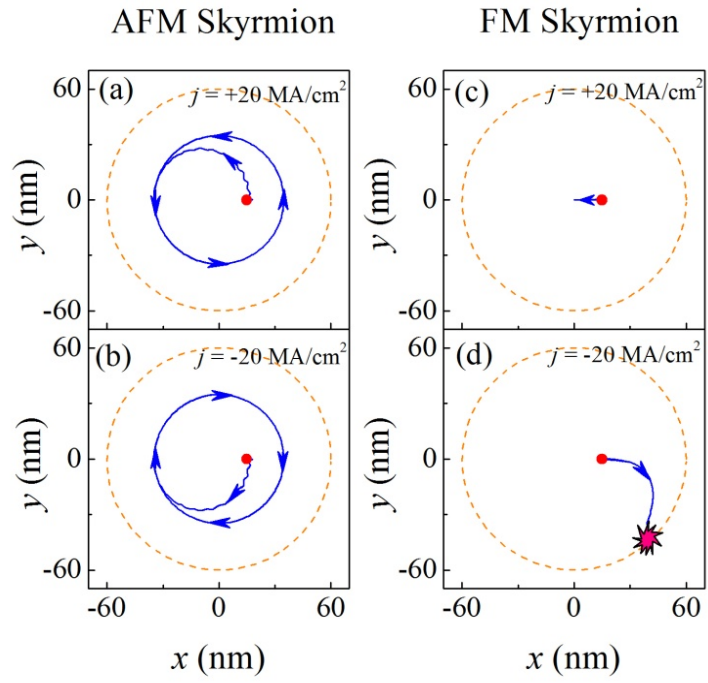

Figure 2

Shen et al.

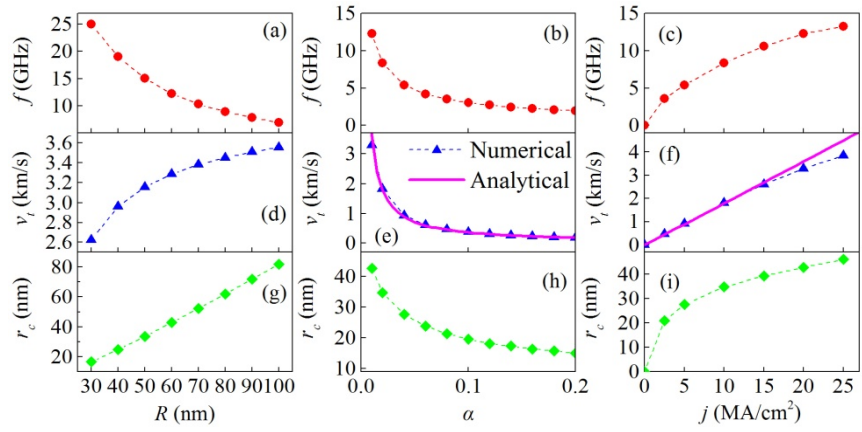

Figure 3

Shen et al.